\documentclass[nofootinbib,aps,showpacs,superscriptaddress,preprintnumbers,epsf,psf]{revtex4-2}
\usepackage[dvips,final]{graphicx}
\usepackage[english]{babel}
\usepackage{pstricks}
\usepackage{epsf}
\usepackage{amsmath}
\usepackage{amsfonts}
\usepackage{amssymb}
\usepackage{dcolumn}
\usepackage{bm}
\usepackage{mathrsfs}
\usepackage{slashed}

\usepackage[colorlinks = true1,
linkcolor = blue,
urlcolor  = blue,
citecolor = blue,
anchorcolor = blue]{hyperref}

\begin{document}
	
	
\title{Novel phase transition at Unruh temperature}

\author{G. Yu. Prokhorov}
\email{prokhorov@theor.jinr.ru}
\affiliation{Joint Institute for Nuclear Research, Joliot-Curie 6, Dubna, 141980, Russia}
\affiliation{NRC Kurchatov Institute, Moscow, Russia}

\author{O. V. Teryaev}
\email{teryaev@jinr.ru}
\affiliation{Joint Institute for Nuclear Research, Joliot-Curie 6, Dubna, 141980, Russia}
\affiliation{NRC Kurchatov Institute, Moscow, Russia}

\author{V. I. Zakharov}
\email{vzakharov@itep.ru}
\affiliation{NRC Kurchatov Institute, Moscow, Russia}
\affiliation{Joint Institute for Nuclear Research, Joliot-Curie 6, Dubna, 141980, Russia}

\begin{abstract} \vspace{0.2 cm}
We consider gas of massless fermions at certain temperature $T$ and acceleration $a$. We find a second order phase transition at temperature $T$
approaching the Unruh temperature
$T_U$.
The implications
for hadronization of the quark-gluon plasma produced
in heavy-ion
collisions (HIC)
and for black-hole physics are discussed.
In particular, this novel phase transition may be associated with thermalization in HIC, indicating its analogy with falling into a black hole.
\end{abstract}

\maketitle

\section{Introduction}
\label{sec intro}

As is argued first in  \cite{Unruh:1976db}
an accelerated observer
perceives
Minkowski vacuum as a thermal bath with temperature
\begin{eqnarray}\label{TU}
T_U=\frac{\hbar|a|}{2 \pi k_B c}\,,
\end{eqnarray}
$|a|=\sqrt{-a^{\mu}a_{\mu}}$, $a_{\mu}=u^{\nu}\partial_{\nu}u_{\mu}$
is
the proper acceleration,
 and $ u_{\mu} $ is four-velocity of the observer\footnote{In what follows we use the system of units $ e=\hbar=c=k_B=1 $.}. This effect is known
as the Unruh effect,
and temperature (\ref{TU}) is Unruh temperature.
One of remarkable features of Unruh temperature is its universality.
It is easiest to derive   (\ref{TU}) in case of noninteracting gases but it
holds in a
generic interacting theory as well \cite{Bisognano:1976za}.

We are primarily interested in the phase diagram in the axes of
acceleration
$ |a|$ and
temperature $T$, as measured
by a comoving observer. Motivation is, to some extent, phenomenological.
As is noted in \cite{STAR:2017ckg, Becattini:2017ljh} the quark-gluon plasma is produced
in heavy-ion
collisions in accelerated and heated state. Moreover, the temperature is
of order of
Unruh temperature \cite{Kharzeev:2005iz} and the process of thermalization of
plasma
has been visualized in \cite{Kharzeev:2005iz} as phase transition to
Hawking-type
radiation,
which could resolve the puzzle of a fast thermalization of the plasma.
We will try to elaborate this picture, explaining, in particular, how
thermal particles can be observed in the laboratory frame,
rather than in accelerated
 frame.

Up to this point, we seem to understand well basic features of
the acceleration-temperature  interplay
as far as temperature
exceeds the Unruh temperature, $T \ge T_U$. In particular, the energy density $\epsilon_{s=1/2}$ of massless fermions with spin $s=1/2$ as a function of temperature and acceleration is well defined and can be estimated within two approaches: geometric (or gravitational) \cite{Dowker:1987pk} and thermodynamic \cite{Prokhorov:2019cik}
\begin{eqnarray}\label{cov}
\epsilon_{s=1/2} \equiv \langle\hat{T}^0_0\rangle = 
\left( \frac{7 \pi^2 T^4}{60}+
\frac{|a|^2 T^2}{24}-\frac{17 |a|^4}{960\pi^2}\right)
\quad (T\ge T_U)\,.
\end{eqnarray}
Note that $\epsilon_{s=1/2}$ vanishes at Unruh temperature, which is a manifestation of subtraction of an ultraviolet divergence, and
the subtraction point
has been fixed as $T=T_U$ on physical grounds. Moreover, such a vanishing at the Unruh temperature
was suggested in \cite{Becattini:2017ljh} as a new definition of Unruh effect \footnote{Undoubtedly, free fermions are just a toy model for a quark-gluon plasma, the use of which is motivated by the fact that quarks can become free due to asymptotic freedom, as well as by the mentioned universality of the Unruh effect.}.

However, for a few reasons we are driven to
extend analysis to lower
temperatures, $T<T_U$, where naive extrapolation of (\ref{cov}) would lead to negative energy. On the phenomenological side,
the plasma produced in heavy-ion collisions  has temperature,
at later times, of about the
Unruh temperature \cite{Kharzeev:2005iz,Castorina:2007eb} while at first moments
after collision $T<T_U$, which is confirmed by direct numerical modeling of HIC  \cite{Shokhonov:prep}. Moreover, for the phase transition of the type discussed in a
\cite{Kharzeev:2005iz,Castorina:2007eb} to
happen one needs initial state with temperature $T<T_U$, which must also be unstable (see below).


To get insight into the region $T < T_U$ we use the geometrical approach
generalizing
the language well known in case $T>T_U$. Namely, we consider massless
fermions living
on the Euclidean version of the Rindler space with a conical singularity.,
see, e.g.,
\cite{Bezerra:2006nu}. The
corresponding
angular deficit is given by
\begin{eqnarray}
\phi_{\text{cone}}~=~2\pi \left(1-\frac{|a|}{2\pi T}\right)\,.
\end{eqnarray}
Now we need to introduce
a negative conical angle deficit $ \phi_{\text{cone}}<0 $ which is to be understood
 as an
analytical continuation (better to say, ``stretching'') of ordinary space,
see, e.g., \cite{Fursaev:1996uz}.
The resulting expression for the energy density differs from
(\ref{cov}) at $T< T_U$ and heat capacity is discontinuous at $T=T_U$.
Moreover, there
are further points of non-analyticity at temperatures $T_n$
\begin{equation}\label{tn}
T_n~=~T_U/(2n+1)\quad (n=0,1,2...)\,.
\end{equation}
Each time, the non-analyticity is associated  with a change in the energy of a single level (more precisely, the two lowest Matsubara modes become singular at the horizon and abruptly change their solutions), which makes this phase transition similar to a quantum phase transition \cite{sachdev_2011}.

We proceed now to details of derivations. They are addressed in the central section \ref{sec mode},
where Matsubara modes in Euclidean Rindler space are used to calculate Green’s functions
and stress-energy tensor. This is followed by the consideration of applications in Section \ref{sec disc}, where falling inside black hole, and hadronization in HIC are considered and compared. In conclusions, Section \ref{sec concl}, we emphasize some specific features of phenomenological description  of processes in accelerated frames.


\section{Singularities of Matsubara modes on the horizon}
\label{sec mode}

\subsection{Euclidean Rindler space: two solutions for the modes}
\label{subsec space}

Medium with a finite proper temperature $ T $ and acceleration
$ a_{\mu} $ (directed along the axis $ z $)
is naturally described as embedded in the Euclidean Rindler space
with the metric
\begin{eqnarray}
ds^2= \frac{\rho^2}{\nu^2} d\varphi^2+d\text{x}^2+d\text{y}^2+
d\rho^2\,,
\label{metricE}
\end{eqnarray}
where $ 0\le\varphi<2 \pi $. The corresponding manifold, $ \mathcal{M} =
\mathbb{R}^2 \otimes \mathcal{C}^2_{\nu} $ ,
contains a 2D cone $\mathcal{C}^2_{\nu}$ with an
angular deficit $2\pi(1-\nu^{-1})$
and is an example of space with a conical singularity, for a review see, e.g.,
\cite{Vilenkin:1984ib} and also \cite{Dowker:1987pk, Fursaev:1996uz}. The $ \varphi $ coordinate corresponds
 to the (periodic and imaginary) proper time $\tau = \rho \varphi/\nu$.
Acceleration and temperature acquire geometric interpretation:
 distance $\rho$ from the apex of the cone to a point on the cone
equals inverse
acceleration $ |a|^{-1} $, and the circumference of a circle with $\rho=const$,
corresponds to the inverse temperature $ T^{-1} $, that is
\begin{eqnarray}
\rho = |a|^{-1} \,,
\quad \nu=\frac{2 \pi T}{|a|}=\text{const}\,. \label{vocab}
\end{eqnarray}
For the metric (\ref{metricE}) the curvature is zero everywhere,
except for the apex of the cone, where it has a delta-function singularity
(which can be ignored for our purposes).
Moreover, the plane $ \rho = 0 $ corresponds to the horizon since
$ g_{00}(\rho = 0)= 0 $.
The region $ T>|a|/2\pi $ (or $ \nu>1 $) has been thoroughly studied, see, e.g.,
\cite{Dowker:1987pk,Frolov:1987dz,
Bezerra:2006nu,Linet:1994tz}, and we will extend  results obtained
to the region $ T<|a|/2\pi $ (or $ \nu<1 $), where the cone angular deficit
becomes negative.

We consider massless Dirac fermions living on the manifold $\mathcal{M}$
with metric (\ref{metricE}).
It is convenient to use a symmetric, Euclidean vierbein of the form
\cite{Linet:1994tz} $ e^{\mu}_{(a)} = \text{diag}(\nu/\rho,1,1,1) $.
The Dirac operator is given by $\slashed{D}_{x} =
\gamma^{\mu}_E\nabla_{\mu}$,
where
$\gamma^{\mu}_E = i^{-1+\delta_{o\mu}}\gamma^{\mu}_{N}$  are curved-space Euclidean gamma matrices,
$\nabla_{\mu}$
are covariant (also  Euclidean) derivatives.
The Green's functions associated with the Dirac operator and its
square are
denoted as $ S_{E}(x;x') $ and $ G_{E}(x;x') $, respectively, where $ x=(\varphi,\text{x},\text{y},\rho) $, and satisfy the
following equations
\begin{eqnarray}
\slashed{D}_{x} S_{E}(x;x')= \slashed{D}_{x}^2 G_{E}(x;x')=
-I_{4} \frac{\delta^{4}(x-x')}{\sqrt{g}}\,,
\label{DD2}
\end{eqnarray}
where $I_4$ is the identity matrix $4\times 4$.
The function $G_{E}(x;x')$ can be constructed from the eigenmodes
$\phi(x)$ of
$ \slashed{D}_{x}^2 $ \cite{Bezerra:2006nu}
\begin{eqnarray}
\slashed{D}_{x}^2 \phi(x) = -\lambda^2 \phi(x)\,,
\label{laplacian proper}
\end{eqnarray}
which are antiperiodic in imaginary time,
$ \phi\left( \varphi+2\pi n \right) = (-1)^n\phi\left(\varphi\right) $,
as it follows from the choice of the vierbein.
Of key importance is that the equation (\ref{laplacian proper}) has two
independent solutions with positive and negative Bessel function index
\begin{eqnarray}
\phi^{\pm}_{q}(x)=\frac{\sqrt{\nu}}{4\pi^{3/2}}\,
e^{i p_{\text{x}}\text{x}+
i p_{\text{y}}\text{y}+i(n+
\frac{1}{2})\varphi} J_{\pm\beta_{s_1}}(\xi \rho)\,
w_{(s_1,s_2)}\,,
\label{two sol}
\end{eqnarray}
where  $s_{1,2}=\pm 1,$~$ \beta_{s_1}=\nu(n+\frac{1}{2})-\frac{s_1}{2} $, $ \xi^2 =
\lambda^2-p_{\text{x}}^2-p_{\text{y}}^2 $,
$ w_{(\pm,+)} =(1,0,\pm 1,0) $ and $ w_{(\pm,-)} =(0,1,0,\mp 1) $,
and $n$ is an integer. Solutions (\ref{two sol}) are classified according to the eigenvalues $ q=(p_{\text{x}},p_{\text{y}},n+1/2,\lambda ,is_1/2,s_2 /2)  $ of the mutually commuting operators.
Eigenfunctions (\ref{two sol}) are actually Matsubara modes, as evidenced, for example, by the fact that the last term in the exponential factor can be written as
$ i(n+\frac{1}{2})\varphi =i\, \pi T(2n+1)\tau $.

Since $ J_a(x)\sim x^a $ at $ x\to 0 $, only one of the two
solutions (\ref{two sol})
for each $ q $ is finite on the horizon, $ \rho\to 0 $.
We impose a standard
  condition
that the modes are to be finite on the horizon.
When passing through $ T=T_U $ we have then to  change the solutions
for the two
lowest Matsubara modes
with $n=0, s_1=1 $ and $ n=-1,s_1=-1 $, so that they
remain finite on the
horizon:
$ \phi^+_{(n=0,\, s_1=1)} ~ \to ~ \phi^-_{(n=0,
\, s_1=1)} $ and  $ \phi^-_{(n=-1,\, s_1=-1)} ~ \to ~ \phi^+_{(n=-1,\, s_1=-1)} $.
Obviously, this situation  repeats itself when we go lower in temperature,
that is, we
have an infinite series of critical points
(\ref{tn}).
At each of these points,
two modes change their form. The choice of the final solution can be fixed by introducing the modulus $|\beta_{s_1}|$ in the index of the Bessel function (\ref{two sol}).

Thus, at the Unruh temperature, the solution for the modes undergoes a restructuring, which already indicates non-analyticity at this point.


\subsection{Green's function and stress-energy tensor}
\label{subsec prop}

Green's function $G_E(x;x')$ can be constructed according to the known rules from the eigenmodes (\ref{two sol}), where it is necessary to choose only a solution that is finite on the horizon.
In the massless case, the final expression in the coordinate representation can be obtained analytically
\begin{eqnarray}
G^E(x;x'|N_0) &=& \frac{\nu \left[\sinh{\left(\frac{1+\nu}{2}\vartheta -
\vartheta \nu N_0\right)}
e^{i(2N_0+1)\Delta\phi \Sigma_0}-\sinh{\left(\frac{1-\nu}{2}\vartheta -
\vartheta \nu N_0\right)}
e^{i(2N_0 -1)\Delta\phi \Sigma_0}\right]}{8\pi^2 \rho\rho'
\sinh{\vartheta}
\left[\cosh{(\nu \vartheta)}-
\cos{\Delta \varphi}\right]}\,,\quad\quad\quad
\label{gen res}
\end{eqnarray}
where $ \Delta x= x-x' $, $ \cosh\vartheta =
(\rho^2+{\rho'}^2+\Delta \text{x}^2+\Delta \text{y}^2)/(2\rho\rho') $, $ \Sigma_0= \frac{i}{4}[\gamma^{E}_{(0)},\gamma^{E}_{(3)}] $, $ \gamma^{E}_{(a)} $ are the Dirac matrices in Euclidean Cartesian coordinates
and $N_0$ is an integer part
\begin{eqnarray}
N_0=\left\lfloor \frac{1}{2\nu} +\frac{1}{2}\right\rfloor =
\left\lfloor \frac{|a|}{4 \pi T} +\frac{1}{2}\right\rfloor \,,
\end{eqnarray}
which is equal to the number of complete rotations per angle
$ 2\pi $ that can be made on the cone
$\mathcal{C}_\nu$. At the same time $ N_0 $ is the number of pairs of Matsubara modes that changed their solutions see the preceding subsection).

Thus, the non-analyticity discovered at the mode level led to jumps in the Green's function.
We are primarily interested in the first critical point $ T_0=|a|/2\pi $. At $T>|a|/2\pi$, when $ N_0=0 $,
Green's function $G^E(x;x'|0)$ actually is given by a known expression,
see
\cite{Linet:1994tz,Bezerra:2006nu,Frolov:1987dz}. But at $|a|/6\pi<T<|a|/2\pi$ we have $ N_0=1 $ and
(\ref{gen res}) leads to a new Green's function.
The function $S_E(x;x'|N_0)=\slashed{D}_{x} G_{E}(x;x'|N_0)$ can be obtained from (\ref{gen res})
using (\ref{DD2}).

Mean value of the stress-energy tensor at a finite temperature is given by the standard expression from the renormalized Green function \cite{Linet:1994tz,Bezerra:2006nu}\footnote{The functions $G_E(x;x'|N_0)$ and $S_E(x;x|N_0)$ (and the corresponding
matrix elements)
are singular at $x\to x'$ and can be renormalized by subtracting their
values at $\nu=1$
(using time variable $ \theta= \varphi/\nu  $), so that
$S^{ren}_E=S_E-S_E^0$.}
\begin{eqnarray}
\langle \hat{T}_{\mu\nu}\rangle =
\frac{i}{4}\lim\limits_{x'\to x} \left(\gamma^N_{\mu}\nabla^{N,x}_{\nu}-
\gamma^N_{\mu}\nabla^{N,x'}_{\nu}+ \mu\leftrightarrow\nu
\right) S^{ren}_E(x;x'|N_0)\,,
\label{set def}
\end{eqnarray}
where Dirac matrices
$\gamma^{\mu}_N =e^{N\,\mu}_{(a)} \gamma^{(a)}_N$.
Using (\ref{gen res}) for $ N_0=1 $, we obtain
\begin{eqnarray} \label{cov1}
T_U/3<T<T_U&:\,\,& \langle \hat{T}^{\alpha}_{\beta}
\rangle =
\left( \frac{127 \pi^2 T^4}{60}-\frac{11|a|^2 T^2}{24}-
\frac{17 |a|^4}{960\pi^2}
\right)
\left(u^{\alpha}u_{\beta}-\frac{1}{3}
\Delta^{\alpha}_{\beta}\right)  \\ \nonumber
&& + \left(\pi|a|T^3-\frac{T|a|^3}
{4\pi}\right)\widetilde{\Delta}^{\alpha}_{\beta}
\,,
\end{eqnarray}
where the projectors
$\Delta^{\alpha}_{\beta} =
\delta^{\alpha}_{\beta}-u^{\alpha}u_{\beta}$ and $
\widetilde{\Delta}^{\alpha}_{\beta}=
\Delta^{\alpha}_{\beta}+\frac{a^{\alpha}a_{\beta}}{|a|^2}
$ on the (hyper)surfaces, orthogonal to $ u_{\mu} $
and $ u_{\mu} $, $ a_{\mu} $
are introduced.

\begin{figure*}[!h]
\begin{minipage}{0.45\textwidth}
\centerline{\includegraphics[width=1\textwidth]{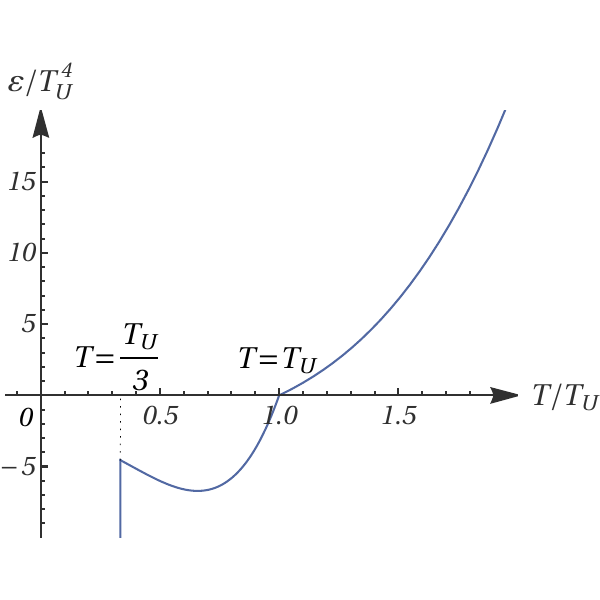}}
\end{minipage}\vspace{1.5cm}
\caption{The energy density as a function of temperature at fixed acceleration. There is a second-order phase transition at the point $ T=T_U $.}
\label{fig:1}
\end{figure*}

Thus, the energy density $ \varepsilon=\langle \hat{T}^{\mu\nu}\rangle u_{\mu}  u_{\nu} $ is different from (\ref{cov}) below the Unruh temperature, although it also becomes negative. Moreover, although the energy density is continuous at the point $T=T_U$, the heat capacity is discontinuous
indicating a second-order phase transition, as it is shown in Figure \ref{fig:1}. Moreover, it turns out that if above the Unruh temperature the trace of the stress-energy tensor is zero, below the Unruh temperature it becomes non-zero
\begin{eqnarray}
T_U/3<T<T_U&:\,\,&\langle  \hat{T}^{\beta}_{\beta}\rangle =
2 \pi T |a| \left(T^2- \frac{|a|^2}{4 \pi^2} \right)\,, \label{trace}
\end{eqnarray}
thus being a possible order parameter.

\section{Applications}
\label{sec disc}

\subsection{Physics of black holes}
\label{subsec bh}

It looks natural to associate the phase diagram in $(T,|a|)$ axes  at
$T<T_U$
with interior region of black holes. Indeed,  from thermodynamical considerations
above
we conclude that at $T<T_U$ the energy  of gas
is negative.   And this matches well
picture that
all the particles inside black holes have
negative
total energy. More precisely, negative modes inside black holes have been considered as a possible source of destruction of the classical manifold, see  \cite{Pimentel:2018fuy}
and references therein.
Also, negative modes in ergosphere of rotating black holes are responsible for the famous
Penrose mechanism of extracting energy from black holes
\cite{Penrose:1964wq}
\footnote{Finally,
let us mention recent analysis  of roles of maximum-entropy
vs minimum-energy
principles in determining ground state of various field theories
\cite{Chandrasekaran:2022cip}.
Interior of
black holes shares many features with De Sitter empty space
which is singled out
by possessing maximal entropy. If this is true also for interior of
black holes then
negative
modes would be natural ``excitations'' of black holes, while
positive-energy modes
would be forbidden.} .

Let us first discuss in more detail the vanishing at the Unruh temperature \cite{Bisognano:1976za, Becattini:2017ljh}. The physically motivated (see below) normalization consists in the assumption that in the Minkowski vacuum state
\begin{eqnarray}\label{minkowski}
T^{\alpha}_{\beta}(\text{Minkowski vacuum})=0 \,.
\end{eqnarray}
Consider now the Minkowski vacuum energy in an accelerated
frame. Since  $T^{\alpha}_{\beta}$ is a tensor with
respect to change of coordinate systems (\ref{minkowski})
remains true in
the accelerated frame as well. On the other hand, according to the Unruh effect, the Minkowski vacuum corresponds to a finite temperature in the accelerated reference frame. Thus, condition (\ref{minkowski}) must be satisfied in any accelerated reference frame at $
T_M = |a_M|/(2\pi)$ (index $ M $ means that the values correspond to the Minkowski vacuum). In other words, choosing an absolute values of $T_M$ (or $ a_M $), is like
fixing a gauge
condition while vanishing (\ref{minkowski}) at the Unruh temperature is
frame-independent
condition.

Let us now turn back to discuss the general constraint $T\ge T_U$ \cite{Becattini:2017ljh}.
This limitation is simply due to the general theorem that the Minkowski vacuum represents
lowest-energy
solution to the Einstein equations \cite{Christodoulou:1993uv}
(for a recent review see \cite{Jia:2023dcu}). In accordance with this, it is physically justified to put all
the components of the
stress-energy tensor equal to zero in the Minkowski vacuum (\ref{minkowski}).
The stability of the Minkowski vacuum assumes that any state has an energy greater than zero and accordingly $ T\ge T_U $.

At the same time, we discovered states $ T<T_U $ for which the energy density is negative, which contradicts stability of the Minkowski vacuum. A possible way out of this situation is precisely the statement that the region $ T<T_U $  refers to the interior of the black hole. Note that despite the negative energy, another criterion for thermodynamic stability is satisfied, namely
\begin{eqnarray}
\frac{d \varepsilon}{d T} > 0\,,
\end{eqnarray}
at least while we are in the region $ T  \lesssim T_U $. Thus, it can be assumed that the matter inside a black hole is stable from this point of view \footnote{Similarly, stability and phase transitions with black holes are discussed in \cite{Witten:2020ert}}.
As far as black hole is stable and interior of it does not
communicate with the
vacuum, the vacuum outside the black hole remains stable. However the existence of negative energies inside a black hole can lead to the formation of thermal Hawking-type radiation, by analogy with \cite{Penrose:1971uk}, which can be achieved through a barrier transition \cite{Morita:2019bfr}
\begin{eqnarray}\label{morita}
(\text{black hole} + \text{matter with}~ T<T_U) ~\to~ \text{(Minkowskian vacuum)} +
\text{(thermal~matter)}\,,
\end{eqnarray}
if such a transition is possible, it will be accomplished through the phase transition at $ T=T_U $ discussed in Section \ref{sec mode}.

In conclusion, we note that presence of negative modes at interior of black holes might allow to avoid application of the Hawking-Penrose theorem \cite{Penrose:1964wq, Hawking:1973uf} on existence of gravitational
singularity inside black holes. Search for ``regular'' black hole solutions is an actual problem, see, e.g., a review \cite{Lan:2023cvz}.

\subsection{Hadronization in quark-gluon plasma}
\label{subsec hadr}

From a phenomenological point of view, when heavy ions collide, the beams interaction region is first de-accelerated as a whole. The potential
energy is accumulated in hadronic strings, connecting quarks. Later, the string
energy is transferred into kinetic, or thermal energy. On the other hand, since the system experiences acceleration, thermalization can be considered as a consequence of the Unruh effect \cite{Kharzeev:2005iz,Castorina:2007eb,Castorina:2018crc}, which allows us to draw an analogy between black holes and confinement. Following this paradigm, we consider the hypothesis that hadronization and thermalization are triggered by acceleration.

However, note that accelerated motion and the Unruh effect by themselves does not automatically lead to particle creation in laboratory frame. A way out of this situation can be found if, in the case of heavy ions, a process similar to (\ref{morita}). Indeed, as already mentioned, at the initial moments of HIC the system is in a state $ T<T_U $ \cite{Shokhonov:prep}, in accordance with the fact that potential energy is accumulated in the strings. The final state is a kinetic-energy dominated, or thermal state, that is $ T>T_U $. These two different states can be distinguished in a frame-independent way.

Since at $ T<T_U $ the energy is less than zero, the corresponding levels can be filled with the formation of positively energetic hadrons, which will be a complete analogy of the process (\ref{morita}). At the same time, as noted in \cite{Morita:2019bfr}, the corresponding spectrum of positive energy particles is thermal and has the Unruh temperature, which is what is needed for thermalization \cite{Kharzeev:2005iz,Castorina:2007eb,Castorina:2018crc}.

The initial state, similar to black hole (\ref{morita}), is disconnected  from the Minkowski vacuum,
providing an escape
from the ``no-go theorem'' \cite{Becattini:2017ljh} .
But unlike the black hole, for which this disconnection takes place in real space, in the case of HIC we assume that disconnection is realized in Hilbert space, when hadronization of the quark-gluon plasma occurs.

This may be also considered as a realization of the picture originally
suggested in \cite{Kharzeev:2005iz} when the statistical approach to Unruh effect
\cite{Becattini:2017ljh, Prokhorov:2019hif} is adopted. The final state thermal hadrons
are produced in the decay of the new phase rather than by the pair creation,
corresponding to standard treatment of Unruh effect.

The qualitative picture of phase transitions in the course of collision of heavy ions may be outlined as follows.
In the initial state, the acceleration dominates over the temperature and hadronic phase persists. The  
hadronic matter is increasing its temperature under the action of (de)accelerating forces and quark-gluon phase may be formed at sufficiently large energies.
At the final state of collision the temperature decreases and the hadrons are formed, which corresponds to ``chemical freeze-out''. The hadrons are then moving with the decreasing acceleration until the ``kinetic freeze-out''.
One should also stress that the very possibility of the existence of temperatures below the  Unruh one is the specific property of hadronic and quark-gluon matter. Say, in the physics at mesoscopic scales the Unruh temperature is much lower than the typical temperature. From this point of view hadronic matter besides being ``extremely hot" is also ``extremely cold".

\section{Conclusions  \& perspectives}
\label{sec concl}

We have investigated the phase diagram in axes of acceleration, $|a|$
 and temperature, $T$. Both $|a|$ and $T$ are universal characteristics
and this universality makes choice of a specific dynamical
model  not crucial.
We consider ideal gas that allows to get explicit answers
to otherwise complicated questions.
For temperature  $T>T_U$  there exists
a well known equilibrium balance between potential energy in
external gravitational
field (or field of  acceleration) and kinetic energy (or temperature).
This balance cannot be extended straightforwardly to  the $T<T_U$
region and,
as a result, one can
suggest  that it is not possible at all to create any substance with $T<T_U$
\cite{Becattini:2017ljh}.

We explore another possibility  and make use of a certain analytical
continuation  to get to $T<T_U$.
Simultaneously one finds
$\varepsilon<0$, where $\varepsilon$ is the total energy density.
The condition $\varepsilon<0$ is apparently realized in interior of
black holes
(and/or their regular modifications, for a review see, e.g.,
\cite{Lan:2023cvz}).
The model might be useful to interpret fastly growing astrophysical evidence on black holes, for overview see, e.g., \cite{Vertogradov:2024seh}.

In a broader context, we are considering physics in non-inertial frames
which is very different
from the common Lorentz invariant case. The main problem  is  that vacuum state of standard 4D field theories is not invariant under coordinate transformations to non-inertial frames so that particle- and vacuum-contributions are mixed.
In particular, in our case
conserved energy includes both energy of
 thermal excitations, and a change in the vacuum energy, or Casimir (or vacuum) effect,
 (see discussion around   (\ref{minkowski})). There is a sum rule
 best known from black-hole physics
\begin{eqnarray}
\epsilon_{\text{thermal}} + \epsilon_{\text{Casimir}}=0 \,.
\end{eqnarray}
In other words, going into accelerated frame
leads to redistribution of thermal and vacuum energies without changing the sum.

To summarize, we have started our discussion by quoting the paper on the Unruh
effect and come out of the discussion with the idea that within thermodynamical approach
there is a kind of unification of the Unruh and Casimir effects which bridges
phenomenology in different fields.

{\bf Acknowledgements}

The authors are thankful to M. Bordag, M. S. Dvornikov,
D. V. Fursaev and
Yu. N. Obukhov for stimulating discussions and comments.

\bibliography{lit}

\begin{thebibliography}{10}

\bibitem{Unruh:1976db}
W.~G. Unruh.
\newblock {Notes on black hole evaporation}.
\newblock {\em Phys. Rev.}, D14:870, 1976.

\bibitem{Bisognano:1976za}
J.~J Bisognano and E.~H. Wichmann.
\newblock {On the Duality Condition for Quantum Fields}.
\newblock {\em J. Math. Phys.}, 17:303--321, 1976.

\bibitem{STAR:2017ckg}
L.~Adamczyk et~al.
\newblock {Global $\Lambda$ hyperon polarization in nuclear collisions:
  evidence for the most vortical fluid}.
\newblock {\em Nature}, 548:62--65, 2017.

\bibitem{Becattini:2017ljh}
F.~Becattini.
\newblock {Thermodynamic equilibrium with acceleration and the Unruh effect}.
\newblock {\em Phys. Rev.}, D97(8):085013, 2018.

\bibitem{Kharzeev:2005iz}
Dmitri Kharzeev and Kirill Tuchin.
\newblock {From color glass condensate to quark gluon plasma through the event
  horizon}.
\newblock {\em Nucl. Phys. A}, 753:316--334, 2005.

\bibitem{Dowker:1987pk}
J.~S. Dowker.
\newblock {Vacuum Averages for Arbitrary Spin Around a Cosmic String}.
\newblock {\em Phys. Rev.}, D36:3742, 1987.

\bibitem{Prokhorov:2019cik}
George~Y. Prokhorov, Oleg~V. Teryaev, and Valentin~I. Zakharov.
\newblock {Unruh effect for fermions from the Zubarev density operator}.
\newblock {\em Phys. Rev.}, D99(7):071901(R), 2019.

\bibitem{Castorina:2007eb}
P.~Castorina, D.~Kharzeev, and H.~Satz.
\newblock {Thermal Hadronization and Hawking-Unruh Radiation in QCD}.
\newblock {\em Eur. Phys. J.}, C52:187--201, 2007.

\bibitem{Shokhonov:prep}
D.~Shohonov~et al.
\newblock In preparation.
\newblock 2024.

\bibitem{Bezerra:2006nu}
Valdir~B. Bezerra and Nail~R. Khusnutdinov.
\newblock {Vacuum expectation value of the spinor massive field in the cosmic
  string space-time}.
\newblock {\em Class. Quant. Grav.}, 23:3449--3462, 2006.

\bibitem{Fursaev:1996uz}
Dmitri~V. Fursaev and Gennaro Miele.
\newblock {Cones, spins and heat kernels}.
\newblock {\em Nucl. Phys. B}, 484:697--723, 1997.

\bibitem{sachdev_2011}
Subir Sachdev.
\newblock {\em Quantum Phase Transitions}.
\newblock Cambridge University Press, 2 edition, 2011.

\bibitem{Vilenkin:1984ib}
Alexander Vilenkin.
\newblock {Cosmic Strings and Domain Walls}.
\newblock {\em Phys. Rept.}, 121:263--315, 1985.

\bibitem{Frolov:1987dz}
Valeri~P. Frolov and E.~M. Serebryanyi.
\newblock {Vacuum Polarization in the Gravitational Field of a Cosmic String}.
\newblock {\em Phys. Rev.}, D35:3779--3782, 1987.

\bibitem{Linet:1994tz}
B.~Linet.
\newblock {Euclidean spinor Green's functions in the space-time of a straight
  cosmic string}.
\newblock {\em J. Math. Phys.}, 36:3694--3703, 1995.

\bibitem{Pimentel:2018fuy}
Guilherme~L. Pimentel, Alexander~M. Polyakov, and Grigory~M. Tarnopolsky.
\newblock {Vacua on the Brink of Decay}.
\newblock pages 399--417, 2018.
\newblock [Rev. Math. Phys.30,no.07,1840013(2018)].

\bibitem{Penrose:1964wq}
Roger Penrose.
\newblock {Gravitational collapse and space-time singularities}.
\newblock {\em Phys. Rev. Lett.}, 14:57--59, 1965.

\bibitem{Chandrasekaran:2022cip}
Venkatesa Chandrasekaran, Roberto Longo, Geoff Penington, and Edward Witten.
\newblock {An algebra of observables for de Sitter space}.
\newblock {\em JHEP}, 02:082, 2023.

\bibitem{Christodoulou:1993uv}
D.~Christodoulou and S.~Klainerman.
\newblock {The Global nonlinear stability of the Minkowski space}.
\newblock 1993.

\bibitem{Jia:2023dcu}
Jin Jia and Pin Yu.
\newblock {Remark on the nonlinear stability of Minkowski spacetime: a rigidity
  theorem}.
\newblock 4 2023.

\bibitem{Witten:2020ert}
Edward Witten.
\newblock {Deformations of JT Gravity and Phase Transitions}.
\newblock 6 2020.

\bibitem{Penrose:1971uk}
R.~Penrose and R.~M. Floyd.
\newblock {Extraction of rotational energy from a black hole}.
\newblock {\em Nature}, 229:177--179, 1971.

\bibitem{Morita:2019bfr}
Takeshi Morita.
\newblock {Thermal Emission from Semi-classical Dynamical Systems}.
\newblock {\em Phys. Rev. Lett.}, 122(10):101603, 2019.

\bibitem{Hawking:1973uf}
Stephen~W. Hawking and George F.~R. Ellis.
\newblock {\em {The Large Scale Structure of Space-Time}}.
\newblock Cambridge Monographs on Mathematical Physics. Cambridge University
  Press, 2 2023.

\bibitem{Lan:2023cvz}
Chen Lan, Hao Yang, Yang Guo, and Yan-Gang Miao.
\newblock {Regular Black Holes: A Short Topic Review}.
\newblock {\em Int. J. Theor. Phys.}, 62(9):202, 2023.

\bibitem{Castorina:2018crc}
Paolo Castorina and Alfredo Iorio.
\newblock {Confinement Horizon and QCD Entropy}.
\newblock {\em Int. J. Mod. Phys. A}, 33(35):1850211, 2018.

\bibitem{Prokhorov:2019hif}
George~Y. Prokhorov, Oleg~V. Teryaev, and Valentin~I. Zakharov.
\newblock {Thermodynamics of accelerated fermion gases and their instability at
  the Unruh temperature}.
\newblock {\em Phys. Rev.}, D100(12):125009, 2019.

\bibitem{Vertogradov:2024seh}
Vitalii Vertogradov and Ali \"Ovg\"un.
\newblock {Exact Regular Black Hole Solutions with de Sitter Cores and Hagedorn
  Fluid}.
\newblock 8 2024.

\end{thebibliography}

\end{document}